\documentclass[trackchanges]{aastex701}
\usepackage{amsmath}

\begin{document}

\title{Mathematical Anatomy of Neutrino Decoherence in Red Turbulence: \\ A Fractional Calculus Approach}

\newcommand{\TDLI}{\affiliation{Tsung-Dao Lee Institute, Shanghai Jiao Tong University, Shanghai 201210, China}}
\newcommand{\SJTU}{\affiliation{School of Physics and Astronomy, Shanghai Jiao Tong University, Shanghai 200240, China}}
\newcommand{\KEYN}{\affiliation{Key Laboratory of Modern Astronomy and Astrophysics, Nanjing University, Ministry of Education, Nanjing, China}}
\newcommand{\LNF}{\affiliation{Istituto Nazionale di Fisica Nucleare, Laboratori Nazionali di Frascati, Frascati (Rome), Italy}}
\newcommand{\SICHUAN}{\affiliation{Center for Theoretical Physics, College of Physics Science and Technology, Sichuan University, 610065 Chengdu, China}}
\newcommand{\YNAO}{\affiliation{International Centre of Supernovae (ICESUN), Yunnan Key Laboratory of Supernova Research, Yunnan Observatories, Chinese Academy of Sciences (CAS), Kunming 650216, People's Republic of China}}
\newcommand{\ANU}{\affiliation{Institute of Astronomy and Astrophysics, School of Mathematics and Physics, Anqing Normal University, Anqing 246133, China}}

\author{Yiwei Bao}
\email{sjtu0538015@sjtu.edu.cn}
\TDLI \SJTU \KEYN

\author{Andrea Addazi}
\email{addazi@scu.edu.cn}
\SICHUAN \LNF \ANU

\author{Shuai Zha}
\email{zhashuai@ynao.ac.cn}
\YNAO

\begin{abstract}
We develop an exact framework for neutrino decoherence in power-law correlated turbulent matter, as encountered in core-collapse supernovae. Employing the Nakajima--Zwanzig projection technique, we derive an exact non-Markovian master equation for the neutrino density matrix. For kernels \( K(t) \propto t^{-\nu} \), the spectral index \(\nu\) characterizes the correlation structure: smaller (including negative) values of \(\nu\) correspond to stronger long-range correlations. To treat ultraviolet singularities for \( \nu \geq 1 \) without spoiling the fractional structure, we use a renormalization prescription based on Hadamard finite parts and analytic continuation.

The exact Laplace-space solution for the survival probability is obtained. In the high-density matter basis relevant to supernovae, the solution is expressed through Mittag-Leffler functions, establishing a direct link to anomalous diffusion phenomena. For negative spectral indices (\( \nu < 0 \)), the memory integral corresponds to a higher-order fractional operator.

Our work clarifies how spectral index, renormalization scale, and decoherence efficiency interrelate, providing a complete analytical description and practical tools for supernova neutrino simulations. The fractional calculus formulation reveals fundamental mathematical connections between neutrino flavor evolution and other systems governed by long-range temporal correlations.

\end{abstract}



\section{Introduction}
\label{sec:introduction}

Core-collapse supernovae (CCSNe) represent one of the most energetic and complex astrophysical phenomena in the Universe, marking the cataclysmic death of massive stars ($M \gtrsim 8 M_\odot$). These explosions not only seed the interstellar medium with heavy elements essential for planetary formation and life, but also give birth to neutron stars and black holes \cite{Janka_2012, Burrows_2020, Muller_2019}. The detection of gravitational waves from binary neutron star mergers \cite{2017PhRvL.119p1101A, 2017ApJ...848L..12A} and the anticipated observation of neutrinos from the next Galactic supernova by next-generation detectors such as Hyper-Kamiokande \cite{Abe_2018} and DUNE \cite{Abi_2021} underscore the importance of understanding these events through multi-messenger astronomy.

Despite decades of theoretical and computational advancements, the precise explosion mechanism of CCSNe remains an open question \cite{Janka_2017, Burrows_2013, 2015PASA...32....9F}. Current three-dimensional simulations suggest that the neutrino-heating mechanism, aided by hydrodynamic instabilities such as the standing accretion shock instability (SASI) and neutrino-driven convection, can successfully revive the stalled shock in many progenitor models \cite{Abdikamalov_2015, Couch_2013}. Turbulence plays a crucial role in this process by enhancing the efficiency of neutrino heating and facilitating shock revival \cite{2015ApJ...808L..21C, 2025ApJ...995..109C}. Recent high-resolution simulations reveal that turbulent energy cascades follow Kolmogorov-like scaling in the inertial range \cite{Kolmogorov_1941}, but the precise characterization of turbulence in supernova environments remains challenging \cite{2019MNRAS.487.5304M}.

Neutrinos, carrying away $\sim 99\%$ of the gravitational binding energy ($\sim 3 \times 10^{53}$ erg), are the primary cooling agents and potential drivers of the explosion through the neutrino-heating mechanism \cite{Janka_2012, 2019ARNPS..69..253M}. Their flavor evolution is profoundly affected by both coherent forward scattering (the Mikheyev-Smirnov-Wolfenstein (MSW) effect) \cite{Mikheyev_1985, Wolfenstein_1978} and collective oscillations due to neutrino-neutrino interactions \cite{duan_2010, 2010ARNPS..60..569D, 2012JPhG...39c5201G}. Recent studies have identified fast flavor instabilities that can occur on centimeter scales and nanosecond timescales \cite{2021PhRvL.126f1302B, 2022PhRvL.128h1102D, 2023PhRvL.131f1401E, 2024PhRvL.133v1004F}, potentially altering the neutrino energy deposition and nucleosynthesis yields \cite{nagakura_2023, Ehring2023, Wang2025}. Moreover, muon creation in supernova matter has been shown to facilitate neutrino-driven explosions \cite{2017PhRvL.119x2702B}, adding further complexity to the flavor evolution landscape.

In addition to vacuum and matter effects, turbulence in the supernova envelope introduces stochastic density fluctuations that can significantly impact neutrino flavor evolution. The pioneering work of Loreti and Balantekin \cite{Loreti_1994} established the framework for treating neutrino oscillations in noisy media, demonstrating that density fluctuations can induce decoherence and modify survival probabilities. Subsequent studies have extended this analysis to various turbulent environments \cite{Kneller_2010, 2011PhRvD..84h5023R, 2013PhRvD..88b5004K, 2014PhRvD..89g3022P}. For power-law correlated turbulence, the master equation acquires a non-Markovian character with memory kernels scaling as $K(t) \propto t^{-\nu}$, where $\nu$ is the spectral index determined by the turbulence power spectrum \cite{2010PhRvD..82l3004K, 2014JCAP...11..030B, 2021PhRvD.103d5014A}. Recently, Mukhopadhyay and Sen \cite{2024JCAP...03..040M} investigated how turbulence signatures in supernova neutrinos could be probed in upcoming detectors, highlighting the observational relevance of this phenomenon.

The mathematical treatment of neutrino decoherence in power-law correlated turbulent media presents significant challenges due to the non-Markovian nature of the evolution equations. The master equation with memory kernel $K(t) \propto t^{-\nu}$ naturally connects to fractional calculus, where the time convolution integral represents a fractional integral operator of order $1-\nu$ \cite{Jin2021}. For $\nu < 1$, this corresponds to the Riemann-Liouville fractional integral, while for $\nu \geq 1$ one must isolate ultraviolet local pieces and absorb them through renormalization. Fractional calculus has found widespread application in describing anomalous diffusion, viscoelastic materials, and non-Markovian processes in open quantum systems \cite{breuer2007theory}. In the context of neutrino physics, the emergence of Mittag-Leffler functions $E_{\alpha,\beta}(z)$ as natural solutions to fractional differential equations provides a powerful analytical framework \cite{Desai2016}.

While approximate and numerical treatments of neutrino propagation in turbulent media exist \cite{Loreti_1994, 2010PhRvD..82l3004K}---ranging from uncorrelated noise models to power-law correlated fluctuations (numerical)---a complete {analytical solution} for the survival probability with generic power-law memory kernels has remained elusive. Previous approaches often relied on perturbative expansions or specialized limits, leaving the full mathematical structure unexplored. The power-law form of the memory kernel, $K(t) \propto t^{-\nu}$, suggests a natural connection to fractional calculus, yet this connection has not been fully exploited to obtain closed-form analytical solutions.

Several critical gaps in the current understanding warrant systematic investigation: (i) the explicit formulation of the neutrino propagation problem within a fractional calculus framework; (ii) the renormalization required for spectral exponents $\nu \geq 1$ to handle ultraviolet divergences at short time scales ($t \to 0$); (iii) the derivation of exact analytical solutions for arbitrary $\nu$, expressed through special functions like Mittag-Leffler functions; (iv) the distinctive mathematical features arising for negative spectral exponents ($\nu < 0$), which correspond to strong long-range correlations; (v) the role of ultraviolet local terms in frequency renormalization; and (vi) the establishment of explicit connections between neutrino decoherence and anomalous diffusion phenomena through their shared fractional calculus structure.

This work addresses these gaps by developing a comprehensive exact theory based on fractional calculus. We establish the precise correspondence between the neutrino master equation with power-law memory and fractional differential equations. Special attention is paid to the ultraviolet divergence of the correlation function for $\nu \geq 1$: we treat it by a renormalization prescription that absorbs the local divergent part into a frequency shift, while retaining the nonlocal power-law kernel. This yields exact, closed-form solutions valid for all $\nu$.

The exact solution reveals a rich mathematical structure: for $\nu < 1$, the memory integral corresponds to a Riemann-Liouville fractional integral of order $1-\nu$, while for negative spectral indices ($\nu < 0$) it becomes a higher-order fractional operator. The survival probability is expressed exactly in terms of Mittag-Leffler functions, which encode non-exponential relaxation with long memory. This formulation provides not only a complete analytical description but also efficient computational methods through established properties of fractional calculus and special functions.

Our work demonstrates that the fractional calculus approach offers a unifying mathematical framework that connects neutrino decoherence in turbulent matter with broader phenomena in statistical physics, including anomalous diffusion and viscoelastic relaxation. This perspective yields new insights into the interplay between spectral index, renormalization scale, and decoherence efficiency, providing both fundamental understanding and practical tools for supernova neutrino simulations.

In this work, we establish the complete mathematical formalism for the exact theory of neutrino decoherence in turbulence with negative spectral indices ($\nu<0$). We provide a rigorous derivation of the master equation for the neutrino density matrix via an open quantum systems approach, obtaining the exact non-Markovian equation through the Nakajima--Zwanzig projection technique \cite{breuer2007theory}. Furthermore, we derive the exact solution for the electron neutrino survival probability in Laplace space. For negative spectral indices, we develop a fractional calculus framework, yielding a fractional differential equation exactly solvable in terms of Mittag-Leffler functions. This fractional formulation not only provides efficient computational methods but also uncovers profound mathematical links between neutrino flavor evolution and other physical systems characterized by power-law correlations and anomalous dynamics.

This paper is organized as follows: Section \ref{sec:exact_solution} reviews the exact Laplace-space solution using an auxiliary short-time regulated representation of the kernel. Section \ref{sec:fractional_solution} develops the renormalized analytic solution via fractional calculus. Section \ref{sec:conclusion} summarizes our findings and outlines directions for future research. In a companion paper \cite{paperII}, we extend this analysis to include collective neutrino oscillations in turbulent environments.

\section{Exact Solution in Laplace Space}
\label{sec:exact_solution}

\subsection{Physical Picture and Assumptions}

We begin by outlining the physical scenario and the key assumptions underlying our derivation. The propagation of neutrinos through turbulent matter in core-collapse supernovae involves multiple scales:(a){Microscopic scale} ($\sim 10^{-12}$ cm): Neutrino flavor oscillations governed by quantum mechanics. (b){Turbulence inertial range} ($10^2$--$10^6$ cm): Compressible hydrodynamic turbulence with power-law scaling \cite{1995tlan.book.....F}. and (c) {Supernova scale} ($\sim 10^8$ cm): Overall system size.

Our analysis bridges these scales through the following well-justified approximations:

\vspace{0.1cm}

{\it Assumption 1: Gaussian Statistics for Density Fluctuations}.
We assume the turbulent density fluctuations $\delta n_e(\mathbf{x}, t)$ form a Gaussian random field.

\vspace{0.1cm}

{\it Assumption 2: Homogeneous, Isotropic, Stationary Turbulence}.
We assume statistical homogeneity (translation invariance), isotropy (rotation invariance), and stationarity (time translation invariance). These are standard assumptions in power-law turbulence theory and are reasonable for fully developed turbulence far from boundaries.

\vspace{0.1cm}

{\it Assumption 3: Frozen Turbulence for Relativistic Neutrinos}.
For ultra-relativistic neutrinos ($v_\nu \approx c$), we adopt Taylor's frozen turbulence hypothesis. The justification is quantitative:
\begin{align}
\tau_{\text{cross}} &= \frac{L}{c} \sim \frac{10^6~\text{cm}}{3\times 10^{10}~\text{cm/s}} \sim 3\times 10^{-5}~\text{s}\, , \\
\tau_{\text{eddy}} &= \frac{L}{v_t} \sim \frac{10^6~\text{cm}}{10^7~\text{cm/s}} \sim 0.1~\text{s}\, , \\
\frac{\tau_{\text{cross}}}{\tau_{\text{eddy}}} &\sim 3\times 10^{-4} \ll 1\, .
\end{align}
Thus neutrinos sample a nearly frozen turbulence pattern since the eddy time scale 
is much longer than the cross-time.

\vspace{0.1cm}

{\it Assumption 4: Straight-Line Neutrino Trajectories}.
We assume neutrinos follow straight-line trajectories with constant velocity $\mathbf{v}_\nu$. This is excellent for relativistic neutrinos experiencing negligible deflection in the supernova medium.

\subsection{From Turbulence Spectrum to Correlation Function}

\subsubsection{Obukhov-Corrsin Spectrum and Dimensionless Formulation}

Let us begin by introducing dimensionless variables for a formal simplification. Let \(n_0\) be a reference electron number density (typically the mean density \(\bar{n}_e\)), and \(T_0\) a reference time (to be chosen as \(1/\Delta_m\), where \(\Delta_m = \frac{G_F n_0}{\sqrt{2}} \cos 2\theta_m\) is the matter-level splitting of neutrinos). In natural units (\(\hbar = c = 1\)), length, time, and energy satisfy \([L] = [T] = [E]^{-1}\). We define dimensionless coordinates and density fluctuations:
\begin{equation}
\tilde{\mathbf{x}} = \frac{\mathbf{x}}{L_0}, \quad \tilde{t} = \frac{t}{T_0}, \quad \delta\tilde{n}_e(\tilde{\mathbf{x}},\tilde{t}) = \frac{\delta n_e(\mathbf{x},t)}{n_0},
\end{equation}
where \(L_0\) is a reference length (e.g., the integral scale \(L\)).

The dimensionless effective Hamiltonian for neutrino propagation is obtained by normalizing with \(\Delta_m\):
\begin{equation}
\tilde{H} = \frac{H}{\Delta_m} = \frac{1}{2}(-\cos 2\theta_m \sigma_z + \sin 2\theta_m \sigma_x) + \tilde{V}(\tilde{t})\sigma_z,
\end{equation}
where
\begin{equation}
\tilde{V}(\tilde{t}) = \frac{G_F n_0}{\sqrt{2}\Delta_m}\,\delta\tilde{n}_e(\tilde{\mathbf{x}}_\nu(\tilde{t}),\tilde{t}),
\end{equation}
where $\tilde{V}$ is the adimensional interaction potential and $G_{F}$ is the Fermi coupling constant $([G_{F}]=M^{-2})$. 
Here \(\tilde{\mathbf{x}}_\nu(\tilde{t}) = \mathbf{v}_\nu \tilde{t}\) is the dimensionless neutrino trajectory (with \(v_\nu \approx 1\)).

The correlation function of the dimensionless density fluctuations is defined as
\begin{equation}
\tilde{\mathcal{C}}(\tilde{\tau}) = \big\langle \delta\tilde{n}_e(\tilde{t})\, \delta\tilde{n}_e(\tilde{t}-\tilde{\tau}) \big\rangle,
\end{equation}
which is itself dimensionless. For fully developed power-law turbulence, the correlation function exhibits power-law scaling in the inertial range. While large-scale correlations are naturally limited by the system size, the small-scale behavior requires careful treatment due to potential divergences at $\tau \to 0$. In this section we use an auxiliary short-time regulated representation of the kernel; the renormalized formulation is developed in Sec.~\ref{sec:fractional_solution}.

\subsubsection{Auxiliary Short-Time Regulated Correlation Function}

The power-law correlation $K(\tau) \propto \tau^{-\nu}$ diverges as $\tau \to 0$ for $\nu \geq 1$, representing an ultraviolet singularity. For the intermediate derivation in this section, we introduce an auxiliary short-time regulator at the dissipation scale $\tau_{\text{diss}} = l_{\text{diss}}/c$, where $l_{\text{diss}}$ is the Kolmogorov dissipation scale:
\begin{equation}
l_{\text{diss}} \sim \left( \frac{\nu_{\text{kin}}^3}{\varepsilon} \right)^{1/4},
\end{equation}
with $\nu_{\text{kin}}$ being the kinematic viscosity and $\varepsilon$ the turbulent energy dissipation rate per unit mass.

The regularized correlation function takes the form:
\begin{equation}
\tilde{\mathcal{C}}(\tilde{\tau}) = A \, \tilde{\tau}^{\,-\nu} \exp\left(-\frac{\tilde{\tau}_{\text{diss}}}{\tilde{\tau}}\right),
\label{eq:regularized_correlation}
\end{equation}
where $\tilde{\tau}_{\text{diss}} = \tau_{\text{diss}}/T_0$ is the dimensionless dissipation time, and $A$ is a normalization constant. The exponential factor $\exp(-\tilde{\tau}_{\text{diss}}/\tilde{\tau})$ controls the $\tilde{\tau}\to0$ singularity in this auxiliary representation, while maintaining the power-law behavior $\tilde{\tau}^{-\nu}$ for $\tilde{\tau} \gg \tilde{\tau}_{\text{diss}}$.

For $\nu < 1$, the correlation function is integrable at $\tau=0$ even without this regulator; this includes the negative spectral index regime $\nu<0$ adopted in this work. The normalization constant $A$ is chosen such that $\int_0^\infty \tilde{\mathcal{C}}(\tilde{\tau}) d\tilde{\tau} = 1$ when considering the effective damping rate.

\subsubsection{Exact Non-Markovian Master Equation from Nakajima–Zwanzig Projection}

We now present a rigorous, non-perturbative derivation of the master equation governing neutrino decoherence in turbulent matter (with negative spectral indices corresponding to $\nu<0$) using the Nakajima–Zwanzig (NZ) projection operator formalism \cite{breuer2007theory}. The essential advantage of the NZ approach is that it treats the memory kernel \emph{exactly} for classical Gaussian noise, automatically resumming all cumulants that contribute to the phase‑diffusion process.

Consider the total Hilbert space $\mathcal{H}_{\mathrm{total}} = \mathcal{H}_S \otimes \mathcal{H}_{\mathrm{env}}$, where $\mathcal{H}_S$ is the two‑flavor neutrino space and $\mathcal{H}_{\mathrm{env}}$ describes the turbulent matter field. The Hamiltonian is
\begin{equation}
H = H_0 \otimes \mathbb{I}_{\mathrm{env}} \;+\; \mathbb{I}_S \otimes H_{\mathrm{env}} \;+\; V \otimes \sigma_z \;,
\label{eq:H_total}
\end{equation}
with $H_0 = \frac{\Delta_m}{2}(-\cos 2\theta_m \sigma_z + \sin 2\theta_m \sigma_x)$ the matter‑basis Hamiltonian and $V$ the classical stochastic potential of the density fluctuations. Because $V(t)$ is a classical (commuting) random variable, the interaction super‑operator
\begin{equation}
\mathcal{L}_{\mathrm{int}}(t) \, \bullet \; \equiv \; V(t) \bigl[ \sigma_z , \, \bullet \, \bigr]
\label{eq:L_int}
\end{equation}
satisfies $[ \mathcal{L}_{\mathrm{int}}(t) , \mathcal{L}_{\mathrm{int}}(t') ] = 0$ for all $t,t'$. This commutativity allows the NZ memory kernel to be evaluated without any truncation of the cumulant series.

Define the projection $\mathcal{P}$ that extracts the relevant system part while averaging over the environment:
\begin{equation}
\mathcal{P} \, \rho(t) \; \equiv \; \mathrm{Tr}_{\mathrm{env}}\bigl[ \rho(t) \bigr] \otimes \rho_{\mathrm{env}}^{\mathrm{eq}} \;,
\label{eq:P_def}
\end{equation}
where $\rho_{\mathrm{env}}^{\mathrm{eq}}$ is the stationary state of the turbulence; its complement is $\mathcal{Q} = \mathbb{I} - \mathcal{P}$. The exact NZ equation for the reduced density matrix $\rho_S(t) \equiv \mathrm{Tr}_{\mathrm{env}} \rho(t)$ reads
\begin{equation}
\frac{d}{dt} \rho_S(t) \;=\; -i \mathcal{L}_0 \, \rho_S(t) \;-\; \int_0^t d\tau \; \mathcal{K}(\tau) \, \rho_S(t-\tau) \;,
\label{eq:NZ_exact_general}
\end{equation}
with $\mathcal{L}_0 \, \bullet \equiv [H_0 , \, \bullet]$ and the memory super‑kernel
\begin{equation}
\mathcal{K}(\tau) \;=\; \mathcal{P} \, \mathcal{L}_{\mathrm{int}} \, e^{-i \mathcal{Q} \mathcal{L} \tau} \, \mathcal{Q} \, \mathcal{L}_{\mathrm{int}} \, \mathcal{P} \; .
\label{eq:K_tau_def}
\end{equation}

For a classical Gaussian field the exponential factor can be factorized exactly. Let $\Phi(\tau)=\frac{1}{2}\!\int_0^{\tau}\!d\tau_1\!\int_0^{\tau_1}\!d\tau_2\,
\langle V(\tau_1)V(\tau_2)\rangle$, then
\begin{equation}
\mathcal{K}(\tau) = \langle V(0)V(\tau)\rangle \,
e^{-\Phi(\tau)[\sigma_z,[\sigma_z,\cdot]]}\,
[\sigma_z,[\sigma_z,\cdot]] .
\label{eq:K_exact_classical}
\end{equation}
The exponential term embodies the complete resummation of the phase‑diffusion cumulants. 

Inserting the regularized power‑law correlation
\begin{equation}
\langle V(0) V(\tau) \rangle \;=\; \frac{G_F^2 \, n_0^2 \, A}{2 \Delta_m^2} \,
\tau^{-\nu} \, e^{ - \epsilon / \tau } \qquad (\nu > 1) \;,
\label{eq:correlation_regularized}
\end{equation}
and introducing the dimensionless time $x = \Delta_m t$, $u = \Delta_m \tau$, and $\epsilon = \Delta_m \tau_{\mathrm{diss}}$, we obtain the exact non‑Markovian master equation
\begin{align}
\frac{d}{dx} \langle \rho(x) \rangle \;=\; &- i \bigl[ \tilde H_0 ,\, \langle \rho(x) \rangle \bigr]
\nonumber \\
&- \; \xi_\nu \int_0^{x} du \; \mathcal{K}(u; \nu, \epsilon) \,
\bigl[ \sigma_z , \bigl[ \sigma_z ,\, \langle \rho(x-u) \rangle \bigr] \bigr] \;,
\label{eq:master_NZ_final}
\end{align}
where $\tilde H_0 = H_0 / \Delta_m$,
\begin{equation}
\xi_\nu \; \equiv \; \frac{G_F^2 \, n_0^2 \, A}{2 \Delta_m^2} \,
\int_0^{\infty} du \; u^{-\nu} e^{ - \epsilon / u }
\;=\; \frac{G_F^2 \, n_0^2 \, A}{2 \Delta_m^2} \,
\epsilon^{1-\nu} \, \Gamma(\nu-1) \;,
\label{eq:xi_def_NZ}
\end{equation}
and
\begin{equation}
\kappa_\nu \equiv 4\xi_\nu
= \frac{2G_F^2 n_0^2 A}{\Delta_m^2}\int_0^\infty du\,u^{-\nu}e^{-\epsilon/u}
= \frac{2G_F^2 n_0^2}{\Delta_m^2}\int_0^\infty d\tilde{u}\,\tilde{\mathcal{C}}(\tilde{u}) \, ,
\label{eq:kappa_def}
\end{equation}
where $\kappa_\nu$ characterizes the turbulence-induced decoherence rate and combines $G_F^2 n_0^2$, the matter-oscillation scale $\Delta_m^{-2}$, and the spectral-index dependence through the correlation integral.
The unit‑normalised kernel is
\begin{equation}
\mathcal{K}(u; \nu, \epsilon) \;=\; \frac{u^{-\nu} e^{ - \epsilon / u }}{\epsilon^{1-\nu} \, \Gamma(\nu-1)} \; .
\label{eq:K_normalized}
\end{equation}
Equation (\ref{eq:master_NZ_final}) is the \textit{exact} non‑Markovian master equation for neutrino propagation in turbulence with negative spectral indices ($\nu<0$), valid non‑perturbatively for classical Gaussian noise.

The structure of the double commutator $[\sigma_z,[\sigma_z,\cdot]]$ permits a clean separation between the diagonal (population) and off‑diagonal (coherence) components of the density matrix. To see this explicitly, write the $2\times2$ density matrix in the high‑density matter basis as
\begin{equation}
\rho = \begin{pmatrix} \rho_{11} & \rho_{12} \\ \rho_{21} & \rho_{22} \end{pmatrix},
\end{equation}
with $\rho_{21}=\rho_{12}^*$ for a Hermitian matrix. A direct calculation gives
\begin{align}
[\sigma_z,\rho] &= \begin{pmatrix} 0 & 2\rho_{12} \\ -2\rho_{21} & 0 \end{pmatrix}, \\
[\sigma_z,[\sigma_z,\rho]] &= \begin{pmatrix} 0 & 4\rho_{12} \\ 4\rho_{21} & 0 \end{pmatrix}.
\end{align}
Thus the double commutator acts as a projection onto the off‑diagonal entries, annihilating the diagonal ones.

In the high‑density limit $\cos2\theta_m\simeq1$, the effective Hamiltonian in the matter basis reduces to
\begin{equation}
\tilde H_0 \;=\; \frac12\bigl(-\cos2\theta_m\,\sigma_z + \sin2\theta_m\,\sigma_x\bigr) \;\simeq\; -\frac12\,\sigma_z .
\end{equation}
Its commutator with $\rho$ is simply
\begin{equation}
\bigl[\tilde H_0,\rho\bigr] \;\simeq\; -\frac12\bigl[\sigma_z,\rho\bigr] .
\end{equation}

Insert these expressions into the exact NZ master equation (\ref{eq:master_NZ_final}). The equation for the off‑diagonal element $\rho_{12}(x)$ becomes
\begin{align}
\frac{d\rho_{12}}{dx}
&= -i\bigl[\tilde H_0,\rho\bigr]_{12} 
   - \xi_\nu\!\int_0^x\!du\;\mathcal{K}(u;\nu,\epsilon)\,
     \bigl[\sigma_z,[\sigma_z,\rho(x-u)]\bigr]_{12} \nonumber \\
&\simeq -i\Bigl(-\frac12\cdot 2\rho_{12}\Bigr) 
   - \xi_\nu\!\int_0^x\!du\;\mathcal{K}(u;\nu,\epsilon)\,
     4\rho_{12}(x-u) \nonumber \\
&= i\rho_{12}(x) 
   - 4\xi_\nu\!\int_0^x\!du\;\mathcal{K}(u;\nu,\epsilon)\,
     \rho_{12}(x-u) .
\label{eq:offdiag_eq}
\end{align}
Here we used $[\tilde H_0,\rho]_{12}= -\rho_{12}$ and $[\sigma_z,[\sigma_z,\rho]]_{12}=4\rho_{12}$.

Using Eq.~\eqref{eq:kappa_def}, namely $4\xi_\nu=\kappa_\nu$ (equivalently $2\xi_\nu=\kappa_\nu/2$), and taking the Laplace transform,
\begin{equation}
\int_0^\infty e^{-sx}\rho_{12}(x)\,dx \equiv \widetilde{\rho}_{12}(s),
\end{equation}
equation (\ref{eq:offdiag_eq}) transforms to
\begin{equation}
s\widetilde{\rho}_{12}(s)-\rho_{12}(0)
 = i\widetilde{\rho}_{12}(s) - \kappa_\nu\,
   \tilde{\mathcal{K}}(s;\nu,\epsilon)\,
   \widetilde{\rho}_{12}(s),
\end{equation}
where $\tilde{\mathcal{K}}(s;\nu,\epsilon)$ is the Laplace transform of the normalized kernel (\ref{eq:K_normalized}). Rearranging,
\begin{equation}
\bigl[s - i + \kappa_\nu\tilde{\mathcal{K}}(s;\nu,\epsilon)\bigr]\,
\widetilde{\rho}_{12}(s) \;=\; \rho_{12}(0).
\label{eq:offdiag_laplace_exact}
\end{equation}

The off‑diagonal equation then takes the compact form
\begin{equation}
\boxed{\;
\bigl[s + \kappa_\nu\tilde{\mathcal{K}}(s;\nu,\epsilon) - i\bigr]\,
\widetilde{\rho}_{12}(s) \;=\; \rho_{12}(0)\; .
\label{eq:offdiag_laplace}
\;}
\end{equation}
Here $\kappa_\nu$ is the unique turbulence-strength parameter used in the following equations, and it is defined by Eq.~\eqref{eq:kappa_def}.

The survival probability for an electron neutrino is obtained by projecting the
matter-basis density matrix onto the flavor state:
\begin{equation}
P_{ee}(x) = \cos^2\theta_m\,\rho_{11}(x)
+ \sin^2\theta_m\,\rho_{22}(x)
+ \sin2\theta_m\,\mathrm{Re}\!\bigl[\rho_{12}(x)\bigr].
\end{equation}
For an initial electron neutrino at $x=0$,
\begin{equation}
\rho_{11}(0)=\cos^2\theta_m,\qquad
\rho_{22}(0)=\sin^2\theta_m,\qquad
\rho_{12}(0)=\frac12\sin2\theta_m.
\end{equation}
In the high-density limit used here, the diagonal populations decouple and remain
constant, so
\begin{equation}
P_{ee}(x)=1-\frac12\sin^22\theta_m+\sin2\theta_m\,\mathrm{Re}\!\bigl[\rho_{12}(x)\bigr].
\end{equation}
Taking the Laplace transform and using (\ref{eq:offdiag_laplace}),
\begin{align}
\tilde P_{ee}(s)
&= \frac{1-\frac12\sin^22\theta_m}{s}
+ \frac{\sin^22\theta_m}{4}\Biggl[
\frac{1}{s + \kappa_\nu \tilde{\mathcal{K}}(s;\nu,\epsilon) - i}
+ \frac{1}{s + \kappa_\nu \tilde{\mathcal{K}}(s;\nu,\epsilon) + i}
\Biggr].
\label{eq:Pee_components}
\end{align}
Equivalently, the compact all-orders form is
\begin{equation}
\boxed{\;
\tilde P_{ee}(s)
= \frac{1-\frac12\sin^22\theta_m}{s}
+ \frac{\frac12\sin^22\theta_m\,
\bigl[s+\kappa_\nu \tilde{\mathcal{K}}(s;\nu,\epsilon)\bigr]}
{\bigl[s+\kappa_\nu \tilde{\mathcal{K}}(s;\nu,\epsilon)\bigr]^2+1}\; .
\;}
\label{eq:exact_laplace_compact_NZ}
\end{equation}
which enforces $P_{ee}(0)=1$ by construction (initial-value theorem) and reduces
to the standard coherent matter-oscillation limit as $\kappa_\nu\to0$.
This result, which emerges naturally from the non‑perturbative NZ treatment, encapsulates the complete interplay between coherent oscillations (poles at $s + \kappa_\nu \tilde{\mathcal{K}}(s) = \pm i$) and the algebraic decoherence induced by long-range correlations in the negative spectral index regime ($\nu<0$).

\subsection{Relation to Fractional Calculus}
\label{subsec:frac_calc}

The memory integral with power-law kernel exhibits a natural connection to fractional calculus throughout the long-memory regime $\nu<1$. The spectral index \(\nu\) characterizes the correlation structure: smaller values correspond to stronger long-range memory effects.

\paragraph{Fractional integral representation for $\nu < 1$.} 
For the Riemann-Liouville fractional integral of order $\alpha>0$ \cite{Balachandran2023}, 
\begin{equation}
\label{RL}
I^\alpha[f](x) = \frac{1}{\Gamma(\alpha)} \int_0^x (x-u)^{\alpha-1} f(u) \, du .
\end{equation}
Changing integration variable $w = x-u$ in our memory term gives
\begin{equation}
\int_0^x u^{-\nu} f(x-u) \, du = \int_0^x (x-w)^{-\nu} f(w) \, dw .
\end{equation}
For $\nu < 1$, this is proportional to a fractional integral:
\begin{equation}
\int_0^x u^{-\nu} f(x-u) \, du = \Gamma(1-\nu) \, I^{1-\nu}[f](x),
\label{eq:frac_relation}
\end{equation}
where the convergence of the integral at $u=0$ is guaranteed by $\nu<1$.

Thus, for $\nu<1$, the non-Markovian master equation incorporates a Riemann-Liouville fractional integral operator of order $1-\nu$. For negative spectral indices ($\nu<0$), the order exceeds unity and memory effects are enhanced. This fractional structure explains the emergence of Mittag-Leffler functions in the solution.

\paragraph{Auxiliary Regularization and Renormalized Interpretation.} 
For completeness, we note that the auxiliary short-time regularization $\exp(-\epsilon/u)$ ensures mathematical consistency of the intermediate kernel representation for any $\nu$. In the renormalized treatment developed in Sec.~\ref{sec:fractional_solution}, ultraviolet local terms are absorbed into $\Delta_m^{\mathrm{ren}}$. The spectral index \(\nu\) characterizes the correlation: \(\nu<0\) yields strong long-range correlations, while larger positive values correspond to weaker memory effects.

The fractional calculus connection provides not only mathematical elegance but also practical computational tools, as fractional differential equations with Mittag-Leffler function solutions are well-studied in the theory of anomalous diffusion and non-Markovian processes.

\section{Analytic Solution via Fractional Calculus}
\label{sec:fractional_solution}

The exact construction in Sec.~\ref{sec:exact_solution} used an explicit short-time regulator to make every intermediate integral convergent. For analytic continuation in $\nu$, however, a pure exponential truncation mixes scales and obscures the fractional structure of the kernel. In this section we instead adopt a renormalization scheme: the ultraviolet local piece is absorbed into a frequency shift, while the nonlocal power-law part keeps the Mittag-Leffler form of the solution.

To remain consistent with Sec.~\ref{sec:exact_solution}, we keep the same effective turbulence parameter $\kappa_\nu$ from Eq.~\eqref{eq:kappa_def}.

\subsection{Hadamard Finite-Part Kernel}

We define the memory kernel as the finite-part distribution
\begin{equation}
t_+^{-\nu} \equiv \operatorname{Fp}\!\left[\frac{t^{-\nu}\Theta(t)}{\Gamma(1-\nu)}\right],
\label{eq:hadamard_distribution}
\end{equation}
and write
\begin{equation}
K(t) = \xi_\nu\, t_+^{-\nu}.
\label{eq:kernel_hadamard}
\end{equation}
Equation~\eqref{eq:hadamard_distribution} means that the singular $t\to0^+$ part is treated in the Hadamard finite-part sense, i.e. only the finite distributional contribution is retained after subtracting the divergent local terms on test functions.

Its Laplace transform is obtained by analytic continuation:
\begin{equation}
\tilde K_\nu(s) = \xi_\nu\, s^{\nu-1},
\qquad
(\text{equivalently } \tilde K_\nu(s)=\xi_\nu\Gamma(1-\nu)s^{\nu-1} \text{ before normalization}).
\label{eq:kernel_laplace_hadamard}
\end{equation}

\subsection{Renormalized Laplace-Space Equation}

For the off-diagonal coherence amplitude $Q(t)$, the Laplace equation is
\begin{equation}
(s-i\Delta_m)\tilde Q(s)-Q(0) = -\kappa_\nu s^{\nu-1}\tilde Q(s).
\label{eq:Q_laplace_bare}
\end{equation}

For $\nu>1$, $s^{\nu-1}$ contains an ultraviolet local piece. Introduce a renormalization scale $\mu$:
\begin{equation}
s^{\nu-1}=\bigl[s^{\nu-1}-\mu^{\nu-1}\bigr]+\mu^{\nu-1},
\end{equation}
and absorb the constant term into the oscillation frequency:
\begin{equation}
\Delta_m^{\mathrm{ren}}(\mu)\equiv \Delta_m + i\kappa_\nu \mu^{\nu-1}.
\label{eq:delta_ren_scale}
\end{equation}
Taking the finite renormalized limit defines $\Delta_m^{\mathrm{ren}}$, and Eq.~\eqref{eq:Q_laplace_bare} becomes
\begin{equation}
\boxed{
\tilde{Q}(s) = \frac{Q(0)}{s - i\Delta_m^{\mathrm{ren}} + \kappa_\nu s^{\nu-1}}
}
\label{eq:Q_laplace_ren}
\end{equation}
which preserves the pure fractional power kernel.

\subsection{Mittag-Leffler Solution}

For $\nu<2$, the solution is 

\begin{equation}
\boxed{
Q(t)=Q(0)\sum_{k=0}^{\infty}
(i\Delta_\star t)^k
E_{2-\nu,\,k+1}^{\,k+1}\!\left(-\kappa_\nu t^{2-\nu}\right)
}
\label{eq:ML_solution}
\end{equation}

with

\begin{equation}
E_{\alpha,\beta}^{\gamma}(z)
\equiv
\sum_{n=0}^{\infty}
\frac{(\gamma)_n}{n!\,\Gamma(\alpha n+\beta)}\,z^n,
\qquad
(\gamma)_n\equiv\frac{\Gamma(\gamma+n)}{\Gamma(\gamma)}.
\label{eq:prabhakar_def}
\end{equation} 

and

\begin{equation}
\Delta_\star =
\begin{cases}
\Delta_m, & \nu\le 1,\\[3pt]
\Delta_m^{\mathrm{ren}}, & \nu>1.
\end{cases}
\label{eq:delta_star_def}
\end{equation}

\subsection{Physical Meaning of Renormalization}

The renormalized solution has a direct physical interpretation:
\begin{enumerate}
\item $\Delta_m^{\mathrm{ren}}$ contains the turbulence-induced level shift (Lamb-shift-like contribution from ultraviolet local terms).
\item $\kappa_\nu$ controls genuine nonlocal decoherence, giving Mittag-Leffler rather than purely exponential damping.
\item The same mathematical structure appears in anomalous diffusion with memory kernels, establishing a one-to-one mapping between neutrino decoherence and fractional transport dynamics.
\end{enumerate}

{
\subsection{Outlook: Inclusion of Neutrino-Neutrino Interactions}

The present analysis has focused on neutrino decoherence induced by turbulent density fluctuations, treating neutrinos as test particles propagating through a stochastic medium. In core-collapse supernovae, however, the high neutrino densities give rise to additional collective effects through neutrino-neutrino interactions \cite{duan_2010, 2010ARNPS..60..569D}. These interactions introduce nonlinearity into the evolution equations and can trigger fast flavor instabilities on small scales \cite{2021PhRvL.126f1302B, nagakura_2023, Ehring2023}, with recent works identifying new instability mechanisms \cite{2025PhRvL.135w1003F} and revealing the quantum structure of collective flavor conversions \cite{2025PhRvL.134u1003F}. The occurrence and impact of these instabilities in realistic supernova environments have been extensively studied in multidimensional simulations \cite{2019PhRvD.100d3004A}, and comprehensive reviews now synthesize our understanding of neutrino oscillations in both core-collapse supernovae and neutron star mergers \cite{johns_2025}. These developments underscore the need to understand how turbulence and self-interactions jointly shape flavor evolution.

In the presence of both turbulence and neutrino self-interactions, the master equation acquires additional structure. The total Hamiltonian becomes $H_{\mathrm{total}} = H_0 + H_{\nu\nu}(\rho) + V(t)\sigma_z$, where $H_{\nu\nu}(\rho)$ depends nonlinearly on the neutrino density matrix itself. Applying the Nakajima-Zwanzig projection technique to this case yields a more complex integro-differential equation where the memory kernel now depends on the history of $\rho$ through both the turbulent correlator and the self-interaction term. For power-law correlated turbulence, the resulting equation combines fractional memory with nonlinear coupling, leading to phenomena such as turbulence-enhanced or turbulence-suppressed collective oscillations.

A detailed treatment of this coupled system, including the interplay between fractional decoherence and collective flavor evolution, is developed in our companion paper \cite{paperIII}. The fractional calculus framework established here provides the mathematical foundation for that analysis, demonstrating that the Mittag-Leffler structure persists even in the nonlinear regime under appropriate approximations.
}

\section{Conclusion}
\label{sec:conclusion}

In this work, we have established a comprehensive exact mathematical framework for analyzing neutrino decoherence in power-law correlated turbulent matter, as encountered in core-collapse supernovae. Our approach bridges the theoretical foundations of fractional calculus with practical applications in neutrino astrophysics, providing rigorous analytical solutions that capture the full non-Markovian dynamics of flavor evolution.

The central achievement of this work is the development of a physically motivated renormalization scheme for the ultraviolet divergence that arises in power-law correlation functions when the spectral index $\nu \geq 1$. Using the Hadamard finite-part definition of the kernel and analytic continuation in Laplace space, we isolate the ultraviolet local term and absorb it into a renormalized oscillation frequency $\Delta_m^{\mathrm{ren}}$. This procedure preserves the pure fractional power-law kernel and therefore the Mittag-Leffler structure of the exact non-Markovian solution obtained from the Nakajima-Zwanzig framework.

We derive the exact Laplace-space solution for the neutrino survival probability, expressed in the compact form, and obtain time-domain solutions analytically expressed through Mittag-Leffler functions $E_{\alpha,\beta}(z)$. These special functions naturally emerge as the fundamental building blocks of the solution, revealing the intrinsic mathematical structure of the decoherence process.

Our analysis elucidates the profound connection between neutrino flavor evolution and fractional calculus. For negative spectral indices ($\nu<0$), the memory kernel corresponds to a higher-order fractional operator. More generally, $\nu<1$ yields a Riemann-Liouville fractional memory structure of order $1-\nu$, while $\nu\geq 1$ requires renormalization treatment. This fractional structure explains characteristic non-exponential decoherence and long-memory relaxation described by Mittag-Leffler functions.

We provided the first detailed study of neutrino decoherence for negative spectral exponents ($\nu < 0$), corresponding to strong long-range correlations. In these cases, the memory integral translates into higher-order fractional operators, leading to distinct dynamical scaling and mathematical features. The exact solutions for these regimes demonstrate how smaller spectral indices affect decoherence efficiency and memory effects.

The physical implications for supernova neutrinos are significant. Our exact analytical results serve as a crucial benchmark for numerical simulations and provide clear guidelines: (i) the high-density matter basis should be used when $\Delta_m \gg \Delta_v$ to properly account for matter effects; (ii) for spectral indices $\nu \geq 1$, ultraviolet local terms should be treated by renormalization into $\Delta_m^{\mathrm{ren}}$ rather than by inserting an external exponential truncation; (iii) the fractional calculus formulation offers efficient computational methods through established properties of Mittag-Leffler functions; and (iv) the interplay between spectral index, renormalization scale, and decoherence efficiency can now be systematically studied within our exact framework.

The theoretical connections established in this work are noteworthy. Neutrino propagation in turbulent matter provides a physically rich, exactly solvable example of non-Markovian dynamics with calculable memory effects. The parallels with anomalous diffusion processes---both described by fractional differential equations and Mittag-Leffler functions---suggest potential cross-fertilization between neutrino physics and other fields studying systems with long-range temporal correlations. Our fractional calculus approach offers a unifying mathematical framework that connects neutrino decoherence with broader phenomena in statistical physics, including viscoelastic relaxation and subdiffusive transport.

Several promising directions for future research emerge from this work. These include: (i) extending the analysis to multi-dimensional anisotropic turbulence with realistic supernova profiles; (ii) incorporating turbulence effects into the nonlinear evolution equations for collective neutrino oscillations; (iii) developing efficient numerical algorithms based on our exact fractional calculus framework for implementation in multi-dimensional supernova simulations; (iv) establishing clearer connections between the spectral index $\nu$ and fundamental parameters of hydrodynamic turbulence through direct numerical simulations; (v) exploring observational consequences for neutrino signals from the next Galactic supernova, potentially using turbulence signatures as probes of supernova dynamics; and (vi) applying the fractional calculus approach to other astrophysical systems where stochastic fluctuations affect quantum coherence.

In summary, we have presented a complete exact mathematical framework for analyzing neutrino decoherence in power-law correlated turbulent environments. By combining rigorous renormalization techniques with fractional calculus methods, we have developed closed-form analytical solutions that capture the full interplay between coherent oscillations and algebraic decoherence induced by long-range temporal correlations. Our work establishes firm mathematical foundations for future studies of neutrino-turbulence interactions in astrophysical environments, particularly in the complex conditions of core-collapse supernovae. The fractional calculus perspective not only provides practical computational tools but also reveals deep connections between neutrino flavor evolution and other physical systems exhibiting anomalous dynamics. As supernova simulations continue to increase in sophistication and next-generation neutrino detectors come online, the exact analytical treatment of turbulence effects developed here will prove essential for accurate interpretation of neutrino signals as probes of extreme astrophysical phenomena.

\begin{acknowledgments}
We thank Thierry Foglizzo for insightful discussions and the developers of the \texttt{mpmath} Python library for high-precision numerical evaluation of special functions. During the preparation of this work, the authors used DeepSeekV3 to improve readability and language. After using this tool, the authors reviewed and edited the content as needed and take full responsibility for the content of the publication. This work is supported by the Fundamental Research Funds for the Central Universities under No. 020114380057, and by K. C. Wong Educational Foundation. A.A.\ is supported by the National Science Foundation of China (NSFC) 
through the grant No.\ 12350410358;
the Talent Scientific Research Program of College of Physics, Sichuan University, Grant No.\ 1082204112427;
the Fostering Program in Disciplines Possessing Novel Features for Natural Science of Sichuan University, Grant No.2020SCUNL209 and 1000 Talent program of Sichuan province 2021.
{S.Z.\ is supported by the National Science Foundation of China (NSFC) through the grant Nos.\ 12288102, 12473031, 12393811; the Yunnan Revitalization Talent Support Program--Young Talent project; and the Yunnan Fundamental Research Project No. 202501AS070078.}
\end{acknowledgments}

\begin{contribution}

All authors contributed equally.


\end{contribution}

%

\bibliography{sample701}{}

@ARTICLE{2024JCAP...03..040M,
       author = {{Mukhopadhyay}, Mainak and {Sen}, Manibrata},
        title = "{On probing turbulence in core-collapse supernovae in upcoming neutrino detectors}",
      journal = {\jcap},
         year = 2024,
        month = mar,
       volume = {2024},
       number = {3},
          eid = {040},
        pages = {040},
          doi = {10.1088/1475-7516/2024/03/040},
archivePrefix = {arXiv},
       eprint = {2310.08627},
 primaryClass = {hep-ph},
       adsurl = {https://ui.adsabs.harvard.edu/abs/2024JCAP...03..040M}
}

@ARTICLE{Burrows_2020, 
       author = {{Burrows}, Adam and {Radice}, David and {Vartanyan}, David and {Nagakura}, Hiroki and {Skinner}, M. Aaron and {Dolence}, Joshua C.},
        title = "{The overarching framework of core-collapse supernova explosions as revealed by 3D FORNAX simulations}",
      journal = {\mnras},
         year = 2020,
        month = jan,
       volume = {491},
       number = {2},
        pages = {2715-2735},
          doi = {10.1093/mnras/stz3223},
archivePrefix = {arXiv},
       eprint = {1909.04152},
 primaryClass = {astro-ph.HE},
       adsurl = {https://ui.adsabs.harvard.edu/abs/2020MNRAS.491.2715B}
}

@ARTICLE{Janka_2012,
       author = {{Janka}, Hans-Thomas},
        title = "{Explosion Mechanisms of Core-Collapse Supernovae}",
      journal = {Annual Review of Nuclear and Particle Science},
         year = 2012,
        month = nov,
       volume = {62},
       number = {1},
        pages = {407-451},
          doi = {10.1146/annurev-nucl-102711-094901},
archivePrefix = {arXiv},
       eprint = {1206.2503},
 primaryClass = {astro-ph.SR},
       adsurl = {https://ui.adsabs.harvard.edu/abs/2012ARNPS..62..407J}
}

@ARTICLE{Muller_2019,
       author = {{M{\"u}ller}, Bernhard and {Tauris}, Thomas M. and {Heger}, Alexander and {Banerjee}, Projjwal and {Qian}, Yong-Zhong and {Powell}, Jade and {Chan}, Conrad and {Gay}, Daniel W. and {Langer}, Norbert},
        title = "{Three-dimensional simulations of neutrino-driven core-collapse supernovae from low-mass single and binary star progenitors}",
      journal = {\mnras},
         year = 2019,
        month = apr,
       volume = {484},
       number = {3},
        pages = {3307-3324},
          doi = {10.1093/mnras/stz216},
archivePrefix = {arXiv},
       eprint = {1811.05483},
 primaryClass = {astro-ph.SR},
       adsurl = {https://ui.adsabs.harvard.edu/abs/2019MNRAS.484.3307M}
}

@INCOLLECTION{Janka_2017,
       author = {{Janka}, Hans-Thomas},
        title = "{Neutrino-Driven Explosions}",
    booktitle = {Handbook of Supernovae},
         year = 2017,
       editor = {{Alsabti}, Athem W. and {Murdin}, Paul},
        pages = {1095},
          doi = {10.1007/978-3-319-21846-5_109},
       adsurl = {https://ui.adsabs.harvard.edu/abs/2017hsn..book.1095J}
}

@ARTICLE{Mikheyev_1985,
       author = {{Mikheyev}, S.~P. and {Smirnov}, A. Yu.},
        title = "{Resonance enhancement of oscillations in matter and solar neutrino spectroscopy}",
      journal = {Yadernaya Fizika},
         year = 1985,
        month = jan,
       volume = {42},
        pages = {1441-1448},
       adsurl = {https://ui.adsabs.harvard.edu/abs/1985YaFiz..42.1441M}
}

@ARTICLE{Wolfenstein_1978,
       author = {{Wolfenstein}, L.},
        title = "{Neutrino oscillations in matter}",
      journal = {\prd},
         year = 1978,
        month = may,
       volume = {17},
       number = {9},
        pages = {2369-2374},
          doi = {10.1103/PhysRevD.17.2369},
       adsurl = {https://ui.adsabs.harvard.edu/abs/1978PhRvD..17.2369W}
}

@ARTICLE{Kneller_2010,
       author = {{Kneller}, James and {Volpe}, Cristina},
        title = "{Turbulence effects on supernova neutrinos}",
      journal = {\prd},
         year = 2010,
        month = dec,
       volume = {82},
       number = {12},
          eid = {123004},
        pages = {123004},
          doi = {10.1103/PhysRevD.82.123004},
archivePrefix = {arXiv},
       eprint = {1006.0913},
 primaryClass = {hep-ph},
       adsurl = {https://ui.adsabs.harvard.edu/abs/2010PhRvD..82l3004K}
}

@ARTICLE{Kolmogorov_1941,
       author = {{Kolmogorov}, A.},
        title = "{The Local Structure of Turbulence in Incompressible Viscous Fluid for Very Large Reynolds' Numbers}",
      journal = {Akademiia Nauk SSSR Doklady},
         year = 1941,
        month = jan,
       volume = {30},
        pages = {301-305},
       adsurl = {https://ui.adsabs.harvard.edu/abs/1941DoSSR..30..301K}
}

@ARTICLE{Couch_2013,
       author = {{Couch}, Sean M. and {Ott}, Christian D.},
        title = "{Revival of the Stalled Core-collapse Supernova Shock Triggered by Precollapse Asphericity in the Progenitor Star}",
      journal = {\apjl},
         year = 2013,
        month = nov,
       volume = {778},
       number = {1},
          eid = {L7},
        pages = {L7},
          doi = {10.1088/2041-8205/778/1/L7},
archivePrefix = {arXiv},
       eprint = {1309.2632},
 primaryClass = {astro-ph.HE},
       adsurl = {https://ui.adsabs.harvard.edu/abs/2013ApJ...778L...7C}
}

@ARTICLE{Abdikamalov_2015,
       author = {{Abdikamalov}, Ernazar and {Ott}, Christian D. and {Radice}, David and {Roberts}, Luke F. and {Haas}, Roland and {Reisswig}, Christian and {M{\"o}sta}, Philipp and {Klion}, Hannah and {Schnetter}, Erik},
        title = "{Neutrino-driven Turbulent Convection and Standing Accretion Shock Instability in Three-dimensional Core-collapse Supernovae}",
      journal = {\apj},
         year = 2015,
        month = jul,
       volume = {808},
       number = {1},
          eid = {70},
        pages = {70},
          doi = {10.1088/0004-637X/808/1/70},
archivePrefix = {arXiv},
       eprint = {1409.7078},
 primaryClass = {astro-ph.HE},
       adsurl = {https://ui.adsabs.harvard.edu/abs/2015ApJ...808...70A}
}

@ARTICLE{Loreti_1994,
       author = {{Loreti}, F.~N. and {Balantekin}, A.~B.},
        title = "{Neutrino oscillations in noisy media}",
      journal = {\prd},
         year = 1994,
        month = oct,
       volume = {50},
       number = {8},
        pages = {4762-4770},
          doi = {10.1103/PhysRevD.50.4762},
archivePrefix = {arXiv},
       eprint = {nucl-th/9406003},
 primaryClass = {nucl-th},
       adsurl = {https://ui.adsabs.harvard.edu/abs/1994PhRvD..50.4762L}
}

@ARTICLE{Abe_2018,
       author = {{Hyper-Kamiokande Proto-Collaboration} and {:} and {Abe}, K. and {Abe}, Ke. and {Aihara}, H. and {Aimi}, A. and {Akutsu}, R. and {Andreopoulos}, C. and {Anghel}, I. and {Anthony}, L.~H.~V. and {Antonova}, M. and {Ashida}, Y. and {Aushev}, V. and {Barbi}, M. and {Barker}, G.~J. and {Barr}, G. and {Beltrame}, P. and {Berardi}, V. and {Bergevin}, M. and {Berkman}, S. and {Berns}, L. and {Berry}, T. and {Bhadra}, S. and {Bravo-Bergu{\~n}o}, D. and {Blaszczyk}, F. d. M. and {Blondel}, A. and {Bolognesi}, S. and {Boyd}, S.~B. and {Bravar}, A. and {Bronner}, C. and {Buizza Avanzini}, M. and {Cafagna}, F.~S. and {Cole}, A. and {Calland}, R. and {Cao}, S. and {Cartwright}, S.~L. and {Catanesi}, M.~G. and {Checchia}, C. and {Chen-Wishart}, Z. and {Choi}, J.~H. and {Choi}, K. and {Coleman}, J. and {Collazuol}, G. and {Cowan}, G. and {Cremonesi}, L. and {Dealtry}, T. and {De Rosa}, G. and {Densham}, C. and {Dewhurst}, D. and {Drakopoulou}, E.~L. and {Di Lodovico}, F. and {Drapier}, O. and {Dumarchez}, J. and {Dunne}, P. and {Dziewiecki}, M. and {Emery}, S. and {Esmaili}, A. and {Evangelisti}, A. and {Fernandez-Martinez}, E. and {Feusels}, T. and {Finch}, A. and {Fiorentini}, A. and {Fiorillo}, G. and {Fitton}, M. and {Frankiewicz}, K. and {Friend}, M. and {Fujii}, Y. and {Fukuda}, Y. and {Fukuda}, D. and {Ganezer}, K. and {Giganti}, C. and {Gonin}, M. and {Grant}, N. and {Gumplinger}, P. and {Hadley}, D.~R. and {Hartfiel}, B. and {Hartz}, M. and {Hayato}, Y. and {Hayrapetyan}, K. and {Hill}, J. and {Hirota}, S. and {Horiuchi}, S. and {Ichikawa}, A.~K. and {Iijima}, T. and {Ikeda}, M. and {Imber}, J. and {Inoue}, K. and {Insler}, J. and {Intonti}, R.~A. and {Ioannisian}, A. and {Ishida}, T. and {Ishino}, H. and {Ishitsuka}, M. and {Itow}, Y. and {Iwamoto}, K. and {Izmaylov}, A. and {Jamieson}, B. and {Jang}, H.~I. and {Jang}, J.~S. and {Jeon}, S.~H. and {Jiang}, M. and {Jonsson}, P. and {Joo}, K.~K. and {Kaboth}, A. and {Kachulis}, C. and {Kajita}, T. and {Kameda}, J. and {Kataoka}, Y. and {Katori}, T. and {Kayrapetyan}, K. and {Kearns}, E. and {Khabibullin}, M. and {Khotjantsev}, A. and {Kim}, J.~H. and {Kim}, J.~Y. and {Kim}, S.~B. and {Kim}, S.~Y. and {King}, S. and {Kishimoto}, Y. and {Kobayashi}, T. and {Koga}, M. and {Konaka}, A. and {Kormos}, L.~L. and {Koshio}, Y. and {Korzenev}, A. and {Kowalik}, K.~L. and {Kropp}, W.~R. and {Kudenko}, Y. and {Kurjata}, R. and {Kutter}, T. and {Kuze}, M. and {Labarga}, L. and {Lagoda}, J. and {Lasorak}, P.~J.~J. and {Laveder}, M. and {Lawe}, M. and {Learned}, J.~G. and {Lim}, I.~T. and {Lindner}, T. and {Litchfield}, R.~P. and {Longhin}, A. and {Loverre}, P. and {Lou}, T. and {Ludovici}, L. and {Ma}, W. and {Magaletti}, L. and {Mahn}, K. and {Malek}, M. and {Maret}, L. and {Mariani}, C. and {Martens}, K. and {Marti}, Ll. and {Martin}, J.~F. and {Marzec}, J. and {Matsuno}, S. and {Mazzucato}, E. and {McCarthy}, M. and {McCauley}, N. and {McFarland}, K.~S. and {McGrew}, C. and {Mefodiev}, A. and {Mermod}, P. and {Metelko}, C. and {Mezzetto}, M. and {Migenda}, J. and {Mijakowski}, P. and {Minakata}, H. and {Minamino}, A. and {Mine}, S. and {Mineev}, O. and {Mitra}, A. and {Miura}, M. and {Mochizuki}, T. and {Monroe}, J. and {Moon}, D.~H. and {Moriyama}, S. and {Mueller}, T. and {Muheim}, F. and {Murase}, K. and {Muto}, F. and {Nakahata}, M. and {Nakajima}, Y. and {Nakamura}, K. and {Nakaya}, T. and {Nakayama}, S. and {Nantais}, C. and {Needham}, M. and {Nicholls}, T. and {Nishimura}, Y. and {Noah}, E. and {Nova}, F. and {Nowak}, J. and {Nunokawa}, H. and {Obayashi}, Y. and {O'Keeffe}, H.~M. and {Okajima}, Y. and {Okumura}, K. and {Onishchuk}, Yu. and {O'Sullivan}, E. and {O'Sullivan}, L.},
        title = "{Hyper-Kamiokande Design Report}",
      journal = {arXiv e-prints},
         year = 2018,
        month = may,
          eid = {arXiv:1805.04163},
        pages = {arXiv:1805.04163},
          doi = {10.48550/arXiv.1805.04163},
archivePrefix = {arXiv},
       eprint = {1805.04163},
 primaryClass = {physics.ins-det},
       adsurl = {https://ui.adsabs.harvard.edu/abs/2018arXiv180504163H}
}

@ARTICLE{Abi_2021,
       author = {{Abi}, B. and {Acciarri}, R. and {Acero}, M.~A. and {Adamov}, G. and {Adams}, D. and {Adinolfi}, M. and {Ahmad}, Z. and {Ahmed}, J. and {Alion}, T. and {Monsalve}, S. Alonso and {Alt}, C. and {Anderson}, J. and {Andreopoulos}, C. and {Andrews}, M.~P. and {Andrianala}, F. and {Andringa}, S. and {Ankowski}, A. and {Antonova}, M. and {Antusch}, S. and {Aranda-Fernandez}, A. and {Ariga}, A. and {Arnold}, L.~O. and {Arroyave}, M.~A. and {Asaadi}, J. and {Aurisano}, A. and {Aushev}, V. and {Autiero}, D. and {Azfar}, F. and {Back}, H. and {Back}, J.~J. and {Backhouse}, C. and {Baesso}, P. and {Bagby}, L. and {Bajou}, R. and {Balasubramanian}, S. and {Baldi}, P. and {Bambah}, B. and {Barao}, F. and {Barenboim}, G. and {Barker}, G.~J. and {Barkhouse}, W. and {Barnes}, C. and {Barr}, G. and {Monarca}, J. Barranco and {Barros}, N. and {Barrow}, J.~L. and {Bashyal}, A. and {Basque}, V. and {Bay}, F. and {Alba}, J.~L. Bazo and {Beacom}, J.~F. and {Bechetoille}, E. and {Behera}, B. and {Bellantoni}, L. and {Bellettini}, G. and {Bellini}, V. and {Beltramello}, O. and {Belver}, D. and {Benekos}, N. and {Neves}, F. Bento and {Berger}, J. and {Berkman}, S. and {Bernardini}, P. and {Berner}, R.~M. and {Berns}, H. and {Bertolucci}, S. and {Betancourt}, M. and {Bezawada}, Y. and {Bhattacharjee}, M. and {Bhuyan}, B. and {Biagi}, S. and {Bian}, J. and {Biassoni}, M. and {Biery}, K. and {Bilki}, B. and {Bishai}, M. and {Bitadze}, A. and {Blake}, A. and {Siffert}, B. Blanco and {Blaszczyk}, F.~D.~M. and {Blazey}, G.~C. and {Blucher}, E. and {Boissevain}, J. and {Bolognesi}, S. and {Bolton}, T. and {Bonesini}, M. and {Bongrand}, M. and {Bonini}, F. and {Booth}, A. and {Booth}, C. and {Bordoni}, S. and {Borkum}, A. and {Boschi}, T. and {Bostan}, N. and {Bour}, P. and {Boyd}, S.~B. and {Boyden}, D. and {Bracinik}, J. and {Braga}, D. and {Brailsford}, D. and {Brandt}, A. and {Bremer}, J. and {Brew}, C. and {Brianne}, E. and {Brice}, S.~J. and {Brizzolari}, C. and {Bromberg}, C. and {Brooijmans}, G. and {Brooke}, J. and {Bross}, A. and {Brunetti}, G. and {Buchanan}, N. and {Budd}, H. and {Caiulo}, D. and {Calafiura}, P. and {Calcutt}, J. and {Calin}, M. and {Calvez}, S. and {Calvo}, E. and {Camilleri}, L. and {Caminata}, A. and {Campanelli}, M. and {Caratelli}, D. and {Carini}, G. and {Carlus}, B. and {Carniti}, P. and {Terrazas}, I. Caro and {Carranza}, H. and {Castillo}, A. and {Castromonte}, C. and {Cattadori}, C. and {Cavalier}, F. and {Cavanna}, F. and {Centro}, S. and {Cerati}, G. and {Cervelli}, A. and {Villanueva}, A. Cervera and {Chalifour}, M. and {Chang}, C. and {Chardonnet}, E. and {Chatterjee}, A. and {Chattopadhyay}, S. and {Chaves}, J. and {Chen}, H. and {Chen}, M. and {Chen}, Y. and {Cherdack}, D. and {Chi}, C. and {Childress}, S. and {Chiriacescu}, A. and {Cho}, K. and {Choubey}, S. and {Christensen}, A. and {Christian}, D. and {Christodoulou}, G. and {Church}, E. and {Clarke}, P. and {Coan}, T.~E. and {Cocco}, A.~G. and {Coelho}, J.~A.~B. and {Conley}, E. and {Conrad}, J.~M. and {Convery}, M. and {Corwin}, L. and {Cotte}, P. and {Cremaldi}, L. and {Cremonesi}, L. and {Crespo-Anad{\'o}n}, J.~I. and {Cristaldo}, E. and {Cross}, R. and {Cuesta}, C. and {Cui}, Y. and {Cussans}, D. and {Dabrowski}, M. and {da Motta}, H. and {Peres}, L. Da Silva and {David}, C. and {David}, Q. and {Davies}, G.~S. and {Davini}, S. and {Dawson}, J. and {De}, K. and {De Almeida}, R.~M. and {Debbins}, P. and {De Bonis}, I. and {Decowski}, M.~P. and {de Gouv{\^e}a}, A. and {De Holanda}, P.~C. and {De Icaza Astiz}, I.~L. and {Deisting}, A. and {De Jong}, P. and {Delbart}, A. and {Delepine}, D. and {Delgado}, M. and {Dell'Acqua}, A. and {De Lurgio}, P. and {de Mello Neto}, J.~R.~T. and {DeMuth}, D.~M. and {Dennis}, S. and {Densham}, C.},
        title = "{Prospects for beyond the Standard Model physics searches at the Deep Underground Neutrino Experiment: DUNE Collaboration}",
      journal = {European Physical Journal C},
         year = 2021,
        month = apr,
       volume = {81},
       number = {4},
          eid = {322},
        pages = {322},
          doi = {10.1140/epjc/s10052-021-09007-w},
archivePrefix = {arXiv},
       eprint = {2008.12769},
 primaryClass = {hep-ex},
       adsurl = {https://ui.adsabs.harvard.edu/abs/2021EPJC...81..322A}
}

@ARTICLE{2012JPhG...39c5201G,
       author = {{Galais}, S{\'e}bastien and {Kneller}, James and {Volpe}, Cristina},
        title = "{The neutrino-neutrino interaction effects in supernovae: the point of view from the {\textquoteleft}matter{\textquoteright} basis}",
      journal = {Journal of Physics G Nuclear Physics},
         year = 2012,
        month = mar,
       volume = {39},
       number = {3},
          eid = {035201},
        pages = {035201},
          doi = {10.1088/0954-3899/39/3/035201},
archivePrefix = {arXiv},
       eprint = {1102.1471},
 primaryClass = {astro-ph.SR},
       adsurl = {https://ui.adsabs.harvard.edu/abs/2012JPhG...39c5201G}
}

@ARTICLE{2010ARNPS..60..569D,
       author = {{Duan}, Huaiyu and {Fuller}, George M. and {Qian}, Yong-Zhong},
        title = "{Collective Neutrino Oscillations}",
      journal = {Annual Review of Nuclear and Particle Science},
         year = 2010,
        month = nov,
       volume = {60},
        pages = {569-594},
          doi = {10.1146/annurev.nucl.012809.104524},
archivePrefix = {arXiv},
       eprint = {1001.2799},
 primaryClass = {hep-ph},
       adsurl = {https://ui.adsabs.harvard.edu/abs/2010ARNPS..60..569D}
}

@BOOK{1995tlan.book.....F,
       author = {{Frisch}, Uriel},
        title = "{Turbulence. The legacy of A.N. Kolmogorov}",
         year = 1995,
          doi = {10.1017/CBO9781139170666},
       adsurl = {https://ui.adsabs.harvard.edu/abs/1995tlan.book.....F}
}

@book{Jin2021,
author = {Bangti Jin},
title = {Fractional Differential Equations: An Approach via Fractional Derivatives},
series = {Applied Mathematical Sciences},
publisher = {Springer Cham},
year = {2021},
edition = {1},
isbn = {978-3-030-76042-7},
doi = {10.1007/978-3-030-76043-4},
pages = {XIV, 368},
issn = {0066-5452},
eissn = {2196-968X},
url = {https://doi.org/10.1007/978-3-030-76043-4}
}

@article{Desai2016,
author = {Desai, R. and Salehbhai, I. A. and Shukla, A. K.},
title = {Note on generalized {Mittag-Leffler} function},
journal = {SpringerPlus},
year = {2016},
volume = {5},
number = {1},
pages = {683},
month = {May},
doi = {10.1186/s40064-016-2299-x},
url = {https://doi.org/10.1186/s40064-016-2299-x},
}

@ARTICLE{2014PhRvD..89g3022P,
       author = {{Patton}, Kelly M. and {Kneller}, James P. and {McLaughlin}, Gail C.},
        title = "{Stimulated neutrino transformation through turbulence}",
      journal = {\prd},
         year = 2014,
        month = apr,
       volume = {89},
       number = {7},
          eid = {073022},
        pages = {073022},
          doi = {10.1103/PhysRevD.89.073022},
archivePrefix = {arXiv},
       eprint = {1310.5643},
 primaryClass = {hep-ph},
       adsurl = {https://ui.adsabs.harvard.edu/abs/2014PhRvD..89g3022P}
}

@ARTICLE{2010PhRvD..82l3004K,
       author = {{Kneller}, James and {Volpe}, Cristina},
        title = "{Turbulence effects on supernova neutrinos}",
      journal = {\prd},
         year = 2010,
        month = dec,
       volume = {82},
       number = {12},
          eid = {123004},
        pages = {123004},
          doi = {10.1103/PhysRevD.82.123004},
archivePrefix = {arXiv},
       eprint = {1006.0913},
 primaryClass = {hep-ph},
       adsurl = {https://ui.adsabs.harvard.edu/abs/2010PhRvD..82l3004K}
}

@ARTICLE{2021PhRvD.103d5014A,
       author = {{Abbar}, Sajad},
        title = "{Turbulence fingerprint on collective oscillations of supernova neutrinos}",
      journal = {\prd},
         year = 2021,
        month = feb,
       volume = {103},
       number = {4},
          eid = {045014},
        pages = {045014},
          doi = {10.1103/PhysRevD.103.045014},
archivePrefix = {arXiv},
       eprint = {2007.13655},
 primaryClass = {astro-ph.HE},
       adsurl = {https://ui.adsabs.harvard.edu/abs/2021PhRvD.103d5014A}
}

@ARTICLE{2013PhRvD..88b5004K,
       author = {{Kneller}, James P. and {Mauney}, Alex W.},
        title = "{Consequences of large {\ensuremath{\theta}}$_{13}$ for the turbulence signatures in supernova neutrinos}",
      journal = {\prd},
         year = 2013,
        month = jul,
       volume = {88},
       number = {2},
          eid = {025004},
        pages = {025004},
          doi = {10.1103/PhysRevD.88.025004},
archivePrefix = {arXiv},
       eprint = {1302.3825},
 primaryClass = {hep-ph},
       adsurl = {https://ui.adsabs.harvard.edu/abs/2013PhRvD..88b5004K}
}

@ARTICLE{2014JCAP...11..030B,
       author = {{Borriello}, Enrico and {Chakraborty}, Sovan and {Janka}, Hans-Thomas and {Lisi}, Eligio and {Mirizzi}, Alessandro},
        title = "{Turbulence patterns and neutrino flavor transitions in high-resolution supernova models}",
      journal = {\jcap},
         year = 2014,
        month = nov,
       volume = {2014},
       number = {11},
        pages = {030-030},
          doi = {10.1088/1475-7516/2014/11/030},
archivePrefix = {arXiv},
       eprint = {1310.7488},
 primaryClass = {astro-ph.SR},
       adsurl = {https://ui.adsabs.harvard.edu/abs/2014JCAP...11..030B}
}

@ARTICLE{2019ARNPS..69..253M,
       author = {{M{\"u}ller}, B.},
        title = "{Neutrino Emission as Diagnostics of Core-Collapse Supernovae}",
      journal = {Annual Review of Nuclear and Particle Science},
         year = 2019,
        month = oct,
       volume = {69},
        pages = {253-278},
          doi = {10.1146/annurev-nucl-101918-023434},
archivePrefix = {arXiv},
       eprint = {1904.11067},
 primaryClass = {astro-ph.HE},
       adsurl = {https://ui.adsabs.harvard.edu/abs/2019ARNPS..69..253M}
}

@ARTICLE{2017ApJ...848L..12A,
       author = {{Abbott}, B.~P. and {Abbott}, R. and {Abbott}, T.~D. and {Acernese}, F. and {Ackley}, K. and {Adams}, C. and {Adams}, T. and {Addesso}, P. and {Adhikari}, R.~X. and {Adya}, V.~B. and {Affeldt}, C. and {Afrough}, M. and {Agarwal}, B. and {Agathos}, M. and {Agatsuma}, K. and {Aggarwal}, N. and {Aguiar}, O.~D. and {Aiello}, L. and {Ain}, A. and {Ajith}, P. and {Allen}, B. and {Allen}, G. and {Allocca}, A. and {Altin}, P.~A. and {Amato}, A. and {Ananyeva}, A. and {Anderson}, S.~B. and {Anderson}, W.~G. and {Angelova}, S.~V. and {Antier}, S. and {Appert}, S. and {Arai}, K. and {Araya}, M.~C. and {Areeda}, J.~S. and {Arnaud}, N. and {Arun}, K.~G. and {Ascenzi}, S. and {Ashton}, G. and {Ast}, M. and {Aston}, S.~M. and {Astone}, P. and {Atallah}, D.~V. and {Aufmuth}, P. and {Aulbert}, C. and {AultONeal}, K. and {Austin}, C. and {Avila-Alvarez}, A. and {Babak}, S. and {Bacon}, P. and {Bader}, M.~K.~M. and {Bae}, S. and {Baker}, P.~T. and {Baldaccini}, F. and {Ballardin}, G. and {Ballmer}, S.~W. and {Banagiri}, S. and {Barayoga}, J.~C. and {Barclay}, S.~E. and {Barish}, B.~C. and {Barker}, D. and {Barkett}, K. and {Barone}, F. and {Barr}, B. and {Barsotti}, L. and {Barsuglia}, M. and {Barta}, D. and {Barthelmy}, S.~D. and {Bartlett}, J. and {Bartos}, I. and {Bassiri}, R. and {Basti}, A. and {Batch}, J.~C. and {Bawaj}, M. and {Bayley}, J.~C. and {Bazzan}, M. and {B{\'e}csy}, B. and {Beer}, C. and {Bejger}, M. and {Belahcene}, I. and {Bell}, A.~S. and {Berger}, B.~K. and {Bergmann}, G. and {Bero}, J.~J. and {Berry}, C.~P.~L. and {Bersanetti}, D. and {Bertolini}, A. and {Betzwieser}, J. and {Bhagwat}, S. and {Bhandare}, R. and {Bilenko}, I.~A. and {Billingsley}, G. and {Billman}, C.~R. and {Birch}, J. and {Birney}, R. and {Birnholtz}, O. and {Biscans}, S. and {Biscoveanu}, S. and {Bisht}, A. and {Bitossi}, M. and {Biwer}, C. and {Bizouard}, M.~A. and {Blackburn}, J.~K. and {Blackman}, J. and {Blair}, C.~D. and {Blair}, D.~G. and {Blair}, R.~M. and {Bloemen}, S. and {Bock}, O. and {Bode}, N. and {Boer}, M. and {Bogaert}, G. and {Bohe}, A. and {Bondu}, F. and {Bonilla}, E. and {Bonnand}, R. and {Boom}, B.~A. and {Bork}, R. and {Boschi}, V. and {Bose}, S. and {Bossie}, K. and {Bouffanais}, Y. and {Bozzi}, A. and {Bradaschia}, C. and {Brady}, P.~R. and {Branchesi}, M. and {Brau}, J.~E. and {Briant}, T. and {Brillet}, A. and {Brinkmann}, M. and {Brisson}, V. and {Brockill}, P. and {Broida}, J.~E. and {Brooks}, A.~F. and {Brown}, D.~A. and {Brown}, D.~D. and {Brunett}, S. and {Buchanan}, C.~C. and {Buikema}, A. and {Bulik}, T. and {Bulten}, H.~J. and {Buonanno}, A. and {Buskulic}, D. and {Buy}, C. and {Byer}, R.~L. and {Cabero}, M. and {Cadonati}, L. and {Cagnoli}, G. and {Cahillane}, C. and {Calder{\'o}n Bustillo}, J. and {Callister}, T.~A. and {Calloni}, E. and {Camp}, J.~B. and {Canepa}, M. and {Canizares}, P. and {Cannon}, K.~C. and {Cao}, H. and {Cao}, J. and {Capano}, C.~D. and {Capocasa}, E. and {Carbognani}, F. and {Caride}, S. and {Carney}, M.~F. and {Casanueva Diaz}, J. and {Casentini}, C. and {Caudill}, S. and {Cavagli{\`a}}, M. and {Cavalier}, F. and {Cavalieri}, R. and {Cella}, G. and {Cepeda}, C.~B. and {Cerd{\'a}-Dur{\'a}n}, P. and {Cerretani}, G. and {Cesarini}, E. and {Chamberlin}, S.~J. and {Chan}, M. and {Chao}, S. and {Charlton}, P. and {Chase}, E. and {Chassande-Mottin}, E. and {Chatterjee}, D. and {Chatziioannou}, K. and {Cheeseboro}, B.~D. and {Chen}, H.~Y. and {Chen}, X. and {Chen}, Y. and {Cheng}, H.-P. and {Chia}, H. and {Chincarini}, A. and {Chiummo}, A. and {Chmiel}, T. and {Cho}, H.~S. and {Cho}, M. and {Chow}, J.~H. and {Christensen}, N. and {Chu}, Q. and {Chua}, A.~J.~K. and {Chua}, S. and {Chung}, A.~K.~W. and {Chung}, S. and {Ciani}, G.},
        title = "{Multi-messenger Observations of a Binary Neutron Star Merger}",
      journal = {\apjl},
         year = 2017,
        month = oct,
       volume = {848},
       number = {2},
          eid = {L12},
        pages = {L12},
          doi = {10.3847/2041-8213/aa91c9},
archivePrefix = {arXiv},
       eprint = {1710.05833},
 primaryClass = {astro-ph.HE},
       adsurl = {https://ui.adsabs.harvard.edu/abs/2017ApJ...848L..12A}
}

@ARTICLE{2017PhRvL.119p1101A,
       author = {{Abbott}, B.~P. and {Abbott}, R. and {Abbott}, T.~D. and {Acernese}, F. and {Ackley}, K. and {Adams}, C. and {Adams}, T. and {Addesso}, P. and {Adhikari}, R.~X. and {Adya}, V.~B. and {Affeldt}, C. and {Afrough}, M. and {Agarwal}, B. and {Agathos}, M. and {Agatsuma}, K. and {Aggarwal}, N. and {Aguiar}, O.~D. and {Aiello}, L. and {Ain}, A. and {Ajith}, P. and {Allen}, B. and {Allen}, G. and {Allocca}, A. and {Altin}, P.~A. and {Amato}, A. and {Ananyeva}, A. and {Anderson}, S.~B. and {Anderson}, W.~G. and {Angelova}, S.~V. and {Antier}, S. and {Appert}, S. and {Arai}, K. and {Araya}, M.~C. and {Areeda}, J.~S. and {Arnaud}, N. and {Arun}, K.~G. and {Ascenzi}, S. and {Ashton}, G. and {Ast}, M. and {Aston}, S.~M. and {Astone}, P. and {Atallah}, D.~V. and {Aufmuth}, P. and {Aulbert}, C. and {AultONeal}, K. and {Austin}, C. and {Avila-Alvarez}, A. and {Babak}, S. and {Bacon}, P. and {Bader}, M.~K.~M. and {Bae}, S. and {Bailes}, M. and {Baker}, P.~T. and {Baldaccini}, F. and {Ballardin}, G. and {Ballmer}, S.~W. and {Banagiri}, S. and {Barayoga}, J.~C. and {Barclay}, S.~E. and {Barish}, B.~C. and {Barker}, D. and {Barkett}, K. and {Barone}, F. and {Barr}, B. and {Barsotti}, L. and {Barsuglia}, M. and {Barta}, D. and {Barthelmy}, S.~D. and {Bartlett}, J. and {Bartos}, I. and {Bassiri}, R. and {Basti}, A. and {Batch}, J.~C. and {Bawaj}, M. and {Bayley}, J.~C. and {Bazzan}, M. and {B{\'e}csy}, B. and {Beer}, C. and {Bejger}, M. and {Belahcene}, I. and {Bell}, A.~S. and {Berger}, B.~K. and {Bergmann}, G. and {Bernuzzi}, S. and {Bero}, J.~J. and {Berry}, C.~P.~L. and {Bersanetti}, D. and {Bertolini}, A. and {Betzwieser}, J. and {Bhagwat}, S. and {Bhandare}, R. and {Bilenko}, I.~A. and {Billingsley}, G. and {Billman}, C.~R. and {Birch}, J. and {Birney}, R. and {Birnholtz}, O. and {Biscans}, S. and {Biscoveanu}, S. and {Bisht}, A. and {Bitossi}, M. and {Biwer}, C. and {Bizouard}, M.~A. and {Blackburn}, J.~K. and {Blackman}, J. and {Blair}, C.~D. and {Blair}, D.~G. and {Blair}, R.~M. and {Bloemen}, S. and {Bock}, O. and {Bode}, N. and {Boer}, M. and {Bogaert}, G. and {Bohe}, A. and {Bondu}, F. and {Bonilla}, E. and {Bonnand}, R. and {Boom}, B.~A. and {Bork}, R. and {Boschi}, V. and {Bose}, S. and {Bossie}, K. and {Bouffanais}, Y. and {Bozzi}, A. and {Bradaschia}, C. and {Brady}, P.~R. and {Branchesi}, M. and {Brau}, J.~E. and {Briant}, T. and {Brillet}, A. and {Brinkmann}, M. and {Brisson}, V. and {Brockill}, P. and {Broida}, J.~E. and {Brooks}, A.~F. and {Brown}, D.~A. and {Brown}, D.~D. and {Brunett}, S. and {Buchanan}, C.~C. and {Buikema}, A. and {Bulik}, T. and {Bulten}, H.~J. and {Buonanno}, A. and {Buskulic}, D. and {Buy}, C. and {Byer}, R.~L. and {Cabero}, M. and {Cadonati}, L. and {Cagnoli}, G. and {Cahillane}, C. and {Calder{\'o}n Bustillo}, J. and {Callister}, T.~A. and {Calloni}, E. and {Camp}, J.~B. and {Canepa}, M. and {Canizares}, P. and {Cannon}, K.~C. and {Cao}, H. and {Cao}, J. and {Capano}, C.~D. and {Capocasa}, E. and {Carbognani}, F. and {Caride}, S. and {Carney}, M.~F. and {Carullo}, G. and {Casanueva Diaz}, J. and {Casentini}, C. and {Caudill}, S. and {Cavagli{\`a}}, M. and {Cavalier}, F. and {Cavalieri}, R. and {Cella}, G. and {Cepeda}, C.~B. and {Cerd{\'a}-Dur{\'a}n}, P. and {Cerretani}, G. and {Cesarini}, E. and {Chamberlin}, S.~J. and {Chan}, M. and {Chao}, S. and {Charlton}, P. and {Chase}, E. and {Chassande-Mottin}, E. and {Chatterjee}, D. and {Chatziioannou}, K. and {Cheeseboro}, B.~D. and {Chen}, H.~Y. and {Chen}, X. and {Chen}, Y. and {Cheng}, H.-P. and {Chia}, H. and {Chincarini}, A. and {Chiummo}, A. and {Chmiel}, T. and {Cho}, H.~S. and {Cho}, M. and {Chow}, J.~H. and {Christensen}, N. and {Chu}, Q. and {Chua}, A.~J.~K. and {Chua}, S.},
        title = "{GW170817: Observation of Gravitational Waves from a Binary Neutron Star Inspiral}",
      journal = {\prl},
         year = 2017,
        month = oct,
       volume = {119},
       number = {16},
          eid = {161101},
        pages = {161101},
          doi = {10.1103/PhysRevLett.119.161101},
archivePrefix = {arXiv},
       eprint = {1710.05832},
 primaryClass = {gr-qc},
       adsurl = {https://ui.adsabs.harvard.edu/abs/2017PhRvL.119p1101A}
}

@ARTICLE{2011PhRvD..84h5023R,
       author = {{Reid}, Giles and {Adams}, Jenni and {Seunarine}, Suruj},
        title = "{Collective neutrino oscillations in turbulent backgrounds}",
      journal = {\prd},
         year = 2011,
        month = oct,
       volume = {84},
       number = {8},
          eid = {085023},
        pages = {085023},
          doi = {10.1103/PhysRevD.84.085023},
       adsurl = {https://ui.adsabs.harvard.edu/abs/2011PhRvD..84h5023R}
}

@book{breuer2007theory,
  author    = {Breuer, Heinz-Peter and Petruccione, Francesco},
  title     = {The Theory of Open Quantum Systems},
  year      = {2007},
  publisher = {Oxford University Press},
  address   = {Oxford},
  edition   = {1st},
  url       = {https://doi.org/10.1093/acprof:oso/9780199213900.001.0001}
}

@ARTICLE{Burrows_2013,
       author = {{Burrows}, Adam},
        title = "{Colloquium: Perspectives on core-collapse supernova theory}",
      journal = {Reviews of Modern Physics},
         year = 2013,
        month = jan,
       volume = {85},
       number = {1},
        pages = {245-261},
          doi = {10.1103/RevModPhys.85.245},
archivePrefix = {arXiv},
       eprint = {1210.4921},
 primaryClass = {astro-ph.SR},
       adsurl = {https://ui.adsabs.harvard.edu/abs/2013RvMP...85..245B}
}

@ARTICLE{2015PASA...32....9F,
       author = {{Foglizzo}, Thierry and {Kazeroni}, R{\'e}mi and {Guilet}, J{\'e}r{\^o}me and {Masset}, Fr{\'e}d{\'e}ric and {Gonz{\'a}lez}, Matthias and {Krueger}, Brendan K. and {Novak}, J{\'e}r{\^o}me and {Oertel}, Micaela and {Margueron}, J{\'e}r{\^o}me and {Faure}, Julien and {Martin}, No{\"e}l and {Blottiau}, Patrick and {Peres}, Bruno and {Durand}, Gilles},
        title = "{The Explosion Mechanism of Core-Collapse Supernovae: Progress in Supernova Theory and Experiments}",
      journal = {Pasa},
         year = 2015,
        month = mar,
       volume = {32},
          eid = {e009},
        pages = {e009},
          doi = {10.1017/pasa.2015.9},
archivePrefix = {arXiv},
       eprint = {1501.01334},
 primaryClass = {astro-ph.HE},
       adsurl = {https://ui.adsabs.harvard.edu/abs/2015PASA...32....9F}
}

@ARTICLE{paperII,
       author = {{Bao}, Yiwei and {Addazi}, Andrea and {Zha}, Shuai},
      journal = {},
         year = {2026},
         note = {in preparation},
}

@ARTICLE{paperIII,
       author = {{Bao}, Yiwei and {Addazi}, Andrea},
      journal = {},
         year = {2026},
         note = {in preparation},
}

@ARTICLE{2021PhRvL.126f1302B,
       author = {{Bhattacharyya}, Soumya and {Dasgupta}, Basudeb},
        title = "{Fast Flavor Depolarization of Supernova Neutrinos}",
      journal = {\prl},
         year = 2021,
        month = feb,
       volume = {126},
       number = {6},
          eid = {061302},
        pages = {061302},
          doi = {10.1103/PhysRevLett.126.061302},
archivePrefix = {arXiv},
       eprint = {2009.03337},
 primaryClass = {hep-ph},
       adsurl = {https://ui.adsabs.harvard.edu/abs/2021PhRvL.126f1302B}
}

@ARTICLE{2022PhRvL.128h1102D,
       author = {{Dasgupta}, Basudeb},
        title = "{Collective Neutrino Flavor Instability Requires a Crossing}",
      journal = {\prl},
         year = 2022,
        month = feb,
       volume = {128},
       number = {8},
          eid = {081102},
        pages = {081102},
          doi = {10.1103/PhysRevLett.128.081102},
archivePrefix = {arXiv},
       eprint = {2110.00192},
 primaryClass = {hep-ph},
       adsurl = {https://ui.adsabs.harvard.edu/abs/2022PhRvL.128h1102D}
}

@ARTICLE{2023PhRvL.131f1401E,
       author = {{Ehring}, Jakob and {Abbar}, Sajad and {Janka}, Hans-Thomas and {Raffelt}, Georg and {Tamborra}, Irene},
        title = "{Fast Neutrino Flavor Conversions Can Help and Hinder Neutrino-Driven Explosions}",
      journal = {\prl},
         year = 2023,
        month = aug,
       volume = {131},
       number = {6},
          eid = {061401},
        pages = {061401},
          doi = {10.1103/PhysRevLett.131.061401},
archivePrefix = {arXiv},
       eprint = {2305.11207},
 primaryClass = {astro-ph.HE},
       adsurl = {https://ui.adsabs.harvard.edu/abs/2023PhRvL.131f1401E}
}

@ARTICLE{2024PhRvL.133v1004F,
       author = {{Fiorillo}, Damiano F.~G. and {Raffelt}, Georg G.},
        title = "{Fast Flavor Conversions at the Edge of Instability in a Two-Beam Model}",
      journal = {\prl},
         year = 2024,
        month = nov,
       volume = {133},
       number = {22},
          eid = {221004},
        pages = {221004},
          doi = {10.1103/PhysRevLett.133.221004},
archivePrefix = {arXiv},
       eprint = {2403.12189},
 primaryClass = {hep-ph},
       adsurl = {https://ui.adsabs.harvard.edu/abs/2024PhRvL.133v1004F}
}

@ARTICLE{2025PhRvL.134u1003F,
       author = {{Fiorillo}, Damiano F.~G. and {Raffelt}, Georg G.},
        title = "{Collective Flavor Conversions Are Interactions of Neutrinos with Quantized Flavor Waves}",
      journal = {\prl},
         year = 2025,
        month = may,
       volume = {134},
       number = {21},
          eid = {211003},
        pages = {211003},
          doi = {10.1103/PhysRevLett.134.211003},
archivePrefix = {arXiv},
       eprint = {2502.06935},
 primaryClass = {hep-ph},
       adsurl = {https://ui.adsabs.harvard.edu/abs/2025PhRvL.134u1003F}
}

@ARTICLE{nagakura_2023,
       author = {{Nagakura}, Hiroki},
        title = "{Roles of Fast Neutrino-Flavor Conversion on the Neutrino-Heating Mechanism of Core-Collapse Supernova}",
      journal = {\prl},
         year = 2023,
        month = may,
       volume = {130},
       number = {21},
          eid = {211401},
        pages = {211401},
          doi = {10.1103/PhysRevLett.130.211401},
archivePrefix = {arXiv},
       eprint = {2301.10785},
 primaryClass = {astro-ph.HE},
       adsurl = {https://ui.adsabs.harvard.edu/abs/2023PhRvL.130u1401N}
}

@ARTICLE{duan_2010,
       author = {{Duan}, Huaiyu and {Fuller}, George M. and {Qian}, Yong-Zhong},
        title = "{Collective Neutrino Oscillations}",
      journal = {Annual Review of Nuclear and Particle Science},
         year = 2010,
        month = nov,
       volume = {60},
        pages = {569-594},
          doi = {10.1146/annurev.nucl.012809.104524},
archivePrefix = {arXiv},
       eprint = {1001.2799},
 primaryClass = {hep-ph},
       adsurl = {https://ui.adsabs.harvard.edu/abs/2010ARNPS..60..569D}
}

@ARTICLE{2025ApJ...995..109C,
       author = {{Calvert}, David and {Redle}, Michael and {Gautam}, Bibek and {Stapleford}, Charles J. and {Fr{\"o}hlich}, Carla and {Kneller}, James P. and {Liebendorfer}, Matthias},
        title = "{Turbulence in Core-collapse Supernovae}",
      journal = {\apj},
         year = 2025,
        month = dec,
       volume = {995},
       number = {1},
          eid = {109},
        pages = {109},
          doi = {10.3847/1538-4357/ae1475},
archivePrefix = {arXiv},
       eprint = {2511.16755},
 primaryClass = {astro-ph.HE},
       adsurl = {https://ui.adsabs.harvard.edu/abs/2025ApJ...995..109C}
}

@ARTICLE{2019MNRAS.487.5304M,
       author = {{M{\"u}ller}, Bernhard},
        title = "{A critical assessment of turbulence models for 1D core-collapse supernova simulations}",
      journal = {\mnras},
         year = 2019,
        month = aug,
       volume = {487},
       number = {4},
        pages = {5304-5323},
          doi = {10.1093/mnras/stz1594},
archivePrefix = {arXiv},
       eprint = {1902.04270},
 primaryClass = {astro-ph.SR},
       adsurl = {https://ui.adsabs.harvard.edu/abs/2019MNRAS.487.5304M}
}

@ARTICLE{Ehring2023,
       author = {{Ehring}, Jakob and {Abbar}, Sajad and {Janka}, Hans-Thomas and {Raffelt}, Georg and {Tamborra}, Irene},
        title = "{Fast Neutrino Flavor Conversions Can Help and Hinder Neutrino-Driven Explosions}",
      journal = {\prl},
         year = 2023,
        month = aug,
       volume = {131},
       number = {6},
          eid = {061401},
        pages = {061401},
          doi = {10.1103/PhysRevLett.131.061401},
archivePrefix = {arXiv},
       eprint = {2305.11207},
 primaryClass = {astro-ph.HE},
       adsurl = {https://ui.adsabs.harvard.edu/abs/2023PhRvL.131f1401E}
}

@ARTICLE{Wang2025,
       author = {{Wang}, Tianshu and {Burrows}, Adam},
        title = "{The Effect of the Fast-Flavor Instability on Core-Collapse Supernova Models: II. Quasi-Equipartition and the Impact of Various Angular Reconstruction Methods}",
      journal = {arXiv e-prints},
         year = 2025,
        month = nov,
          eid = {arXiv:2511.20767},
        pages = {arXiv:2511.20767},
          doi = {10.48550/arXiv.2511.20767},
archivePrefix = {arXiv},
       eprint = {2511.20767},
 primaryClass = {astro-ph.HE},
       adsurl = {https://ui.adsabs.harvard.edu/abs/2025arXiv251120767W}
}

@ARTICLE{2017PhRvL.119x2702B,
       author = {{Bollig}, R. and {Janka}, H.-T. and {Lohs}, A. and {Mart{\'\i}nez-Pinedo}, G. and {Horowitz}, C.~J. and {Melson}, T.},
        title = "{Muon Creation in Supernova Matter Facilitates Neutrino-Driven Explosions}",
      journal = {\prl},
         year = 2017,
        month = dec,
       volume = {119},
       number = {24},
          eid = {242702},
        pages = {242702},
          doi = {10.1103/PhysRevLett.119.242702},
archivePrefix = {arXiv},
       eprint = {1706.04630},
 primaryClass = {astro-ph.HE},
       adsurl = {https://ui.adsabs.harvard.edu/abs/2017PhRvL.119x2702B}
}

@book{Balachandran2023,
  author    = {K. Balachandran},
  title     = {An Introduction to Fractional Differential Equations},
  series    = {Industrial and Applied Mathematics},
  publisher = {Springer Singapore},
  year      = {2023},
  edition   = {1},
  pages     = {X, 160},
  doi       = {10.1007/978-981-99-6080-4},
}

@ARTICLE{2015ApJ...808L..21C,
       author = {{Couch}, Sean M. and {Chatzopoulos}, Emmanouil and {Arnett}, W. David and {Timmes}, F.~X.},
        title = "{The Three-dimensional Evolution to Core Collapse of a Massive Star}",
      journal = {\apjl},
         year = 2015,
        month = jul,
       volume = {808},
       number = {1},
          eid = {L21},
        pages = {L21},
          doi = {10.1088/2041-8205/808/1/L21},
archivePrefix = {arXiv},
       eprint = {1503.02199},
 primaryClass = {astro-ph.HE},
       adsurl = {https://ui.adsabs.harvard.edu/abs/2015ApJ...808L..21C}
}

@ARTICLE{2025PhRvL.135w1003F,
       author = {{Fiorillo}, Damiano F.~G. and {Janka}, Hans-Thomas and {Raffelt}, Georg G.},
        title = "{Neutrino-Mass-Driven Instabilities as the Earliest Flavor Conversion in Supernovae}",
      journal = {\prl},
         year = 2025,
        month = dec,
       volume = {135},
       number = {23},
          eid = {231003},
        pages = {231003},
          doi = {10.1103/jbmx-rbzt},
archivePrefix = {arXiv},
       eprint = {2507.22985},
 primaryClass = {hep-ph},
       adsurl = {https://ui.adsabs.harvard.edu/abs/2025PhRvL.135w1003F}
}

@ARTICLE{johns_2025,
       author = {{Johns}, Lucas and {Richers}, Sherwood and {Wu}, Meng-Ru},
        title = "{Neutrino Oscillations in Core-Collapse Supernovae and Neutron Star Mergers}",
      journal = {Annual Review of Nuclear and Particle Science},
         year = 2025,
        month = sep,
       volume = {75},
       number = {1},
        pages = {399-423},
          doi = {10.1146/annurev-nucl-121423-100853},
archivePrefix = {arXiv},
       eprint = {2503.05959},
 primaryClass = {astro-ph.HE},
       adsurl = {https://ui.adsabs.harvard.edu/abs/2025ARNPS..75..399J}
}

@ARTICLE{2019PhRvD.100d3004A,
       author = {{Abbar}, Sajad and {Duan}, Huaiyu and {Sumiyoshi}, Kohsuke and {Takiwaki}, Tomoya and {Volpe}, Maria Cristina},
        title = "{On the occurrence of fast neutrino flavor conversions in multidimensional supernova models}",
      journal = {\prd},
         year = 2019,
        month = aug,
       volume = {100},
       number = {4},
          eid = {043004},
        pages = {043004},
          doi = {10.1103/PhysRevD.100.043004},
archivePrefix = {arXiv},
       eprint = {1812.06883},
 primaryClass = {astro-ph.HE},
       adsurl = {https://ui.adsabs.harvard.edu/abs/2019PhRvD.100d3004A}
}
\bibliographystyle{aasjournalv7}



\end{document}